\documentclass[twocolumn,showpacs,preprintnumbers,amsmath,amssymb, showkeys]{revtex4}
\usepackage{graphicx}% Include figure files
\usepackage{dcolumn}% Align table columns on decimal point
\usepackage{bm}% bold math
\usepackage{epsf}

\begin{document}

%\preprint{APS/123-QED}

\title{Charge trapping in the system of interacting quantum dots}

\author{V.\,N.\,Mantsevich}
 \altaffiliation{vmantsev@spmlab.phys.msu.ru}
\author{N.\,S.\,Maslova}%
 \email{spm@spmlab.phys.msu.ru}
\affiliation{%
 Moscow State University, Department of  Physics,
119991 Moscow, Russia
}%
\author{P.\,I.\,Arseyev}%
 \email{ars@lpi.ru}
\affiliation{%
 Lebedev Institute of Physics, RAS,
119991 Moscow, Russia }

\date{\today }
7 pages, 4 figures
\begin{abstract}
We analyzed the localized charge dynamics in the system of $N$
interacting single-level quantum dots (QDs) coupled to the
continuous spectrum states in the presence of Coulomb interaction
between electrons within the dots. Different dots geometry and
initial charge configurations were considered. The analysis was
performed by means of Heisenberg equations for localized electrons
pair correlators.

We revealed that charge trapping takes place for a wide range of
system parameters and we suggested the QDs geometry for experimental
observations of this phenomenon. We demonstrated significant
suppression of Coulomb correlations with the increasing of QDs
number. We found the appearance of several time scales with the
strongly different relaxation rates for a wide range of the Coulomb
interaction values.

\end{abstract}

\maketitle

\section{Introduction}
Coupled quantum dots (QDs) are recently under numerous experimental
\cite{Wiel},\cite{Potok},\cite{Hayashi} and theoretical
investigations
\cite{Stafford},\cite{Matveev},\cite{Boese},\cite{Kikoin0},\cite{Orellana},\cite{Arseyev_1},\cite{Arseyev_2}
due to their potential application in modern nanoscale devices
dealing with quantum kinetics of individual localized states
\cite{Arseyev_1},\cite{Arseyev_2},\cite{Stafford_1},\cite{Hazelzet},\cite{Cota},\cite{Contreras-Pulido},\cite{Elste},\cite{Kennes}.
The kinetic properties of coupled QDs (artificial molecules)
\cite{Wiel} are governed by the Coulomb interaction between the
localized electrons \cite{Hayashi},\cite{Arseyev_2} and depend
strongly on the dots topology, which determines energy levels
spacing and the coupling rates
\cite{Kastner},\cite{Beenakker},\cite{Alhassid}. During the last
decade experimental technique gives possibility to create vertically
aligned strongly interacting QDs with only one of them coupled to
the continuous spectrum states \cite{Vamivakas},\cite{Stinaff}. This
so-called side-coupled geometry gives an opportunity to analyze
non-stationary effects in formation of various charge and spin
configurations in the small size structures
\cite{Kikoin0},\cite{Arseyev_2}. Lateral QDs are extremely tunable
by means of individual electrical gates
\cite{Kastner_1},\cite{Ashoori}. This advantage reveals in the
possibility of single electron localization in the system of several
coupled dots \cite{Chan} and charge states manipulations in the
artificial molecules. Therefore lateral QDs are considered to be an
ideal candidates for creation of an efficient charge traps. Previous
studies demonstrated long-lived charge occupation trap states in a
single QDs \cite{Kuno_1},\cite{Kuno_2},\cite{Hummon} and single
electron spin trapping \cite{Brum}. Single electron trapping in the
double dot system was performed in \cite{Pioro}. The temperature of
the trapped electron was measured and tunnel coupling energy was
extracted by charge sensing measurements. A full
configuration-interaction study on a square QD containing several
electrons in the presence of an attractive impurity was performed in
\cite{Pujari}. Authors demonstrated that the impurity changes
significantly the charge densities of the two-electron QD excited
states. The effect of correlations was revealed in the enhancement
of the charge densities localization within the dot. QDs were
investigated theoretically by various methods such as Keldysh
non-equilibrium Green-function formalism \cite{Keldysh},
re-normalization group theory \cite{Kikoin}, specific approach
suggested by Coleman \cite{Coleman}, spin-density-functional theory
\cite{Reimann} or quantum Monte-Carlo calculations \cite{Foulkes}.

In this paper we consider charge relaxation in the system of $N$
interacting QDs with on-site Coulomb repulsion coupled to the
reservoir (continuous spectrum states). The analysis was performed
by means of Heisenberg equations for the localized electrons pair
correlators. We demonstrated the presence of strong charge trapping
effects for the lateral QDs geometry. We found that on-site Coulomb
repulsion results in the significant changing of the localized
charge relaxation and leads to the formation of several time ranges
with strongly different values of the relaxation rates. We also
pointed out the significant suppression of Coulomb correlations
influence on the localized charge relaxation with the increasing of
dots number.

\section{Theoretical model}

Let us consider relaxation processes in the system of $N$ identical
lateral QDs which are situated in the different space points and are
connected only with the single-central QD by means of electron
tunneling processes with the same tunneling transfer amplitudes $T$.
We assume that the single particle level spacing in the dots is
large than all other energy scales, so that only one spin-degenerate
level within the QD spectrum is accessible ($\varepsilon_0$ in the
$N$ identical dots and $\varepsilon$ in the central one). QD with
energy level $\varepsilon$ is also connected with the continuous
spectrum states. Moreover we take into account Coulomb interaction
between the localized electrons within the dots ($U$-in the central
one and $U_{0}$ in the surrounding dots). Hamiltonian of the system
under investigation has the form:

\begin{eqnarray}
\Hat{H}&=&\varepsilon
a_{\sigma}^{+}a_{\sigma}+\sum_{\sigma,j=1}^{N}\varepsilon_{0}b_{j\sigma}^{+}b_{j\sigma}+\sum_{\sigma,j=1}^{N}(Tb_{j\sigma}^{+}a_{\sigma}+Ta_{\sigma}^{+}b_{j\sigma})+\nonumber\\
&+&\varepsilon_{k}c_{k\sigma}^{+}c_{k\sigma}+\sum_{k,\sigma}T_{k}(c_{k\sigma}^{+}a_{\sigma}+a_{\sigma}^{+}c_{k\sigma})+\nonumber\\
&+&Un_{a\sigma}n_{a-\sigma}+U_{0}n_{j\sigma}n_{j-\sigma}\
\label{hamiltonian}
\end{eqnarray}

$T_{k}$ - tunneling amplitude between the single-central dot and
continuous spectrum states. We assume $T$ and $T_{k}$ to be
independent of momentum and spin. By considering a constant density
of states in the reservoir $\nu_{0}$ (which is not a function of
energy), the tunneling coupling strength $\gamma$ is defined as
$\gamma=\pi\nu_{0}T_{k}^{2}$.

$a_{\sigma}^{+}/a_{\sigma}$($b_{j\sigma}^{+}/b_{j\sigma}$)-
electrons creation/annihilation operators in the central dot (in the
$N$ surrounding QDs). $c_{k}^{+}/c_{k}$- electrons
creation/annihilation operators in the continuous spectrum states
($k$) and $n_{a\sigma}(n_{j\sigma})$ are electron filling numbers in
the dots.

We'll at first analyze filling numbers relaxation processes in the
case when on-site Coulomb repulsion is absent in the whole system
($U=U_{0}=0$). Let us assume that at the initial moment all charge
density in the system is localized only in one of the $N$ QDs and
has the value $n_{1}(0)=n_{0}$. The filling numbers time evolution
can be analyzed by means of kinetic equations for bilinear
combinations of Heisenberg operators $a_{\sigma}^{+}/a_{\sigma}$ and
$b_{j\sigma}^{+}/b_{j\sigma}$:

\begin{eqnarray}
b_{j\sigma}^{+}b_{j^{'}\sigma}=G_{jj^{'}}(t);a_{\sigma}^{+}a_{\sigma}=G_{aa}(t)\nonumber\\
b_{j\sigma}^{+}a_{\sigma}=G_{ja}(t);a_{\sigma}^{+}b_{j\sigma}=G_{aj}(t)\
\end{eqnarray}

Localized charge time evolution can be obtained from the system of
equations for the Green functions $G_{jj^{'}}$, $G_{aj}$, $G_{ja}$
and $G_{aa}$:

\begin{eqnarray}
i\frac{\partial}{\partial
t}G_{jj^{'}}&=&T\cdot G_{ja}-T\cdot G_{aj^{'}}\nonumber\\
i\frac{\partial}{\partial
t}G_{aa}&=&\sum_{j^{'}}(T\cdot G_{aj^{'}}-T\cdot G_{j^{'}a})-i2\gamma\cdot G_{aa}\nonumber\\
i\frac{\partial}{\partial t}G_{aj}&=&-\Delta\cdot
G_{aj}+T\cdot G_{aa}-\sum_{j^{'}}T\cdot G_{j^{'}j}-i\gamma\cdot G_{aj}\nonumber\\
i\frac{\partial}{\partial t}G_{ja}&=&\Delta\cdot G_{ja}-T\cdot
G_{aa}+\sum_{j^{'}}T\cdot G_{jj^{'}}-i\gamma\cdot G_{ja}\nonumber\\
\label{system_0}
\end{eqnarray}

where $\Delta=\varepsilon-\varepsilon_0$ is the detuning between the
energy levels in the dots.

System of equations (\ref{system_0}) can be re-written in the
compact matrix form (symbol $[]$ means commutation):

\begin{eqnarray}
i\frac{\partial}{\partial
t}\widehat{G}=[\widehat{G},\widehat{A}]-i(\widehat{B}\widehat{G}+\widehat{G}\widehat{B})
\label{system_compact}
\end{eqnarray}

where $\widehat{G}$ is the pair correlators matrix:

\begin{eqnarray}
\hat{G}=
\begin{pmatrix}
G_{aa} & G_{a1} & \ldots & G_{aN}\\
G_{1a} & \ldots & \ldots & G_{1N}\\
\vdots & \vdots & \vdots & \vdots\\
G_{Na} & G_{N1} & \ldots & G_{NN}\\
\end{pmatrix}
\label{matrix}
\end{eqnarray}

and matrix $\widehat{A}$ has the following form:

\begin{eqnarray}
\hat{A}=
\begin{pmatrix}
\Delta & T_{1} & \ldots & T_{N}\\
T_{1}^{*} & 0 & \ldots & 0\\
\vdots & 0 & \ldots & 0\\
T_{N}^{*} & 0 & \ldots & 0\\
\end{pmatrix}
\label{matrix1}
\end{eqnarray}

The tunneling coupling matrix $\widehat{B}$ has only one nonzero
element $||B||_{11}=\gamma$.

The formal solution of the system (\ref{system_compact}) can be
found with the help of evolution operator:

\begin{eqnarray}
\widehat{G}(t)=e^{[{-\widehat{B}t+i\widehat{A}t}]}\widehat{G}(0)
e^{[{\widehat{B}t-i\widehat{A}t}]}
\end{eqnarray}

Consequently the average value of filling numbers time evolution in
the one of the $N$ QDs can be found from the following expression:

\begin{eqnarray}
\langle
n_{j}(t)\rangle=\widehat{G}_{jj}(t)=\sum_{a,b}[e^{{-\widehat{B}t+i\widehat{A}t}}]_{ja}\widehat{G}_{ab}(0)
[e^{{\widehat{B}t-i\widehat{A}t}}]_{bj}
\end{eqnarray}

Due to the condition that initial charge is localized only in one QD
with number $j$, the following initial conditions are fulfilled:
$<n_{j\sigma}(0)>=G_{jj}(0)=n_{0}$, $<n_{a\sigma}(0)>=0$,
$<n_{j^{'}\sigma}(0)>=0$, if $j\neq j^{'}$ and
$G_{jj^{'}}(0)=G_{aj}(0)=G_{ja}(0)=0$.

Let us analyze filling numbers time evolution in the central QD and
in the dot with initial charge. Concerning initial conditions one
can easily find the expressions for filling numbers relaxation:

\begin{eqnarray}
n_{j\sigma}(t)&=&[e^{i\widehat{H}t}]_{jj}n_{0j}[e^{-i\widehat{\widetilde{H}}t}]_{jj}\nonumber\\
n_{a\sigma}(t)&=&[e^{i\widehat{H}t}]_{aj}n_{0j}[e^{-i\widehat{\widetilde{H}}t}]_{ja}\
\label{exponents}
\end{eqnarray}

where operators $\widehat{H}=\widehat{A}+i\widehat{B}$ and
$\widehat{\widetilde{H}}=\widehat{A}-i\widehat{B}$ are included.
Further analysis deals with the calculation of matrix exponent's
elements. One can easily perform this procedure with the help of
recurrent ratio similar to the procedure suggested by Cummings
\cite{Cummings}. The following ratios for operator $\widehat{H}$
elements
 are fulfilled:

\begin{eqnarray}
(\widehat{H}^{n})_{jj}&=&\widehat{H}_{ja}(\widehat{H}^{n-1})_{aj}\nonumber\\
(\widehat{H}^{n})_{aj}&=&\sum_{j^{'}}\widehat{H}_{aj^{'}}(\widehat{H}^{n-1})_{j^{'}j}+\widehat{H}_{aa}(\widehat{H}^{n-1})_{aj}\nonumber\\
(\widehat{H}^{n})_{jj^{'}}&=&T_{j^{'}}^{*}(\widehat{H}^{n-1})_{aj}\
\label{recurent}
\end{eqnarray}

System of equations (\ref{recurent}) enables to get recurrent ratio
for matrix elements $(\widehat{H}^{n})_{aj}$:

\begin{eqnarray}
(\widehat{H}^{n})_{aj}=(\Delta+i\gamma)(\widehat{H}^{n-1})_{aj}+N|T|^{2}(\widehat{H}^{n-2})_{aj}
\end{eqnarray}

analogous equations can be obtained for matrix elements
$(\widehat{\widetilde{H}}^{n})_{aj}$. Consequently after some
calculations one can get:

\begin{eqnarray}
(\widehat{H}^{n})_{aj}&=&\frac{T_{j}}{2\sqrt{D}}(a^{n}+b^{n})\nonumber\\
(\widehat{\widetilde{H}}^{n})_{aj}&=&\frac{T_{j}^{*}}{2\sqrt{\widetilde{D}}}(\widetilde{a}^{n}+\widetilde{b}^{n})\
\label{equation}
\end{eqnarray}

Where coefficients $D$,$\widetilde{D}$,$a$,$\widetilde{a}$,$b$ and
$\widetilde{b}$ are determined as:

\begin{eqnarray}
D=\sqrt{\frac{(\Delta+i\gamma)^{2}}{4}+N|T|^{2}}\nonumber\\
\widetilde{D}=\sqrt{\frac{(\Delta-i\gamma)^{2}}{4}+N|T|^{2}}\nonumber\\
a=\frac{\Delta+i\gamma}{2}+D;
\widetilde{a}=\frac{\Delta-i\gamma}{2}+\widetilde{D}\nonumber\\
b=\frac{\Delta+i\gamma}{2}-D;
\widetilde{b}=\frac{\Delta-i\gamma}{2}-\widetilde{D}\ \label{0}
\end{eqnarray}

Expanding exponents in the expression (\ref{exponents}) in a power
$\widehat{H}$ and $\widehat{\widetilde{H}}$ series one can easily
obtain the following expressions:

\begin{eqnarray}
[e^{i\widehat{H}t}]_{jj}&=&\frac{|T|^{2}}{2D}\cdot[\frac{e^{iat}}{a}-\frac{e^{ibt}}{b}-(\frac{1}{a}-\frac{1}{b})]+1\nonumber\\
e^{[-i\widehat{\widetilde{H}}t]_{jj}}&=&\frac{|T|^{2}}{2\widetilde{D}}\cdot[\frac{e^{-i\widetilde{a}t}}{\widetilde{a}}-\frac{e^{-i\widetilde{b}t}}{\widetilde{b}}-(\frac{1}{\widetilde{a}}-\frac{1}{\widetilde{b}})]+1\
\label{exponents_1}
\end{eqnarray}

After substituting (\ref{exponents_1}) to equations
(\ref{exponents}) one gets expressions which describe filling
numbers time evolution in the central QD $n_{a}(t)$ and in the QD
with the initial charge $n_{j}(t)$ in the case when Coulomb
correlations are neglected.

\begin{eqnarray}
n_{a\sigma}(t)&=&\frac{T^{2}}{4D\widetilde{D}}\cdot(e^{iat}-e^{ibt})\cdot(e^{-i\widetilde{a}t}-e^{-i\widetilde{b}t})\nonumber\\
n_{j\sigma}(t)&=&n_{0j}\cdot[1-\frac{1}{N}+\frac{|T|^{2}}{2D}(\frac{e^{iat}}{a}-\frac{e^{ibt}}{b})]\cdot\nonumber\\&\cdot&[1-\frac{1}{N}+\frac{|T|^{2}}{2\widetilde{D}}(\frac{e^{-i\widetilde{a}t}}{\widetilde{a}}-\frac{e^{-i\widetilde{b}t}}{\widetilde{b}})]\
\label{results}
\end{eqnarray}

It is clearly evident that with the increasing of QDs number $N$
initial charge $n_{0j}$ is quite fully confined in the initial QD
even in the presence of dissipation in the system due to the
interaction with the reservoir.

\begin{eqnarray}
lim_{t\rightarrow\infty}n_{j\sigma}(t)=n_{0j}(1-\frac{1}{N})^{2}\
\end{eqnarray}

Simultaneously for the large number of QDs $N$, electron filling
numbers in the QDs reveal oscillations frequency increasing as
$T\sqrt{N}$. If initial charge is localized in the central QD, which
is coupled to the continuous spectrum states, one should solve
system (\ref{system_0}) with the initial conditions:
$<n_{j\sigma}(0)>=0$, $<n_{a\sigma}(0)>=G_{aa}(0)=n_{0}$,
$<n_{j^{'}\sigma}(0)>=0$. Consequently one can get the following
expressions for the charge time evolution:

\begin{eqnarray}
n_{a\sigma}(t)=[e^{i\widehat{H}t}]_{aa}n_{0a}[e^{-i\widehat{\widetilde{H}}t}]_{aa}\nonumber\\
n_{j\sigma}(t)=[e^{i\widehat{H}t}]_{ja}n_{0a}[e^{-i\widehat{\widetilde{H}}t}]_{aj}\
\label{exp}
\end{eqnarray}

The function $\psi=e^{[i\widehat{H}t]_{aa}}$ can be obtained from
equation:

\begin{eqnarray}
\frac{\partial\psi}{\partial t}=i(\Delta+i\gamma)\cdot\psi+NT\cdot
[e^{i\widehat{H}t}]_{ja}
\end{eqnarray}

Finally, solution will have the form:

\begin{eqnarray}
%\psi=e^{i(\Delta+i\gamma)t}+\int_{0}^{t}NT\cdot e^{i(\Delta+i\gamma)(t-t_{1})}e^{[i\widehat{H}t_{1}]_{ja}}dt_{1}\nonumber\\
\psi=\frac{NT^{2}}{2D}(\frac{e^{iat}-e^{i(\Delta+i\gamma)t}}{a-(\Delta+i\gamma)}-\frac{e^{ibt}-e^{i(\Delta+i\gamma)t}}{b-(\Delta+i\gamma)})
\end{eqnarray}

where coefficients $a$, $b$ and $D$ are determined by the
expressions (\ref{0}). Consequently, the charge trapping effect is
absent in this situation.

We now consider the situation when Coulomb interaction between
localized electrons exists within the QDs. In this case it is
necessary to take into account the following interaction part of the
system Hamiltonian (\ref{hamiltonian}):

\begin{eqnarray}
H_{int}=U_{(0)}n_{\alpha\sigma}n_{\alpha-\sigma}
\end{eqnarray}

where index $\alpha=a(j)$ and Coulomb interaction values $U_{(0)}$
correspond to the central dot(surrounding dots). We'll take into
account Coulomb interaction by means of self-consistent mean field
approximation \cite{Arseyev_2}. It means that the initial energy
level value $\varepsilon$ have to be substituted by the value
$\widetilde{\varepsilon}=\varepsilon+U\cdot<n_{\alpha\sigma}(t)>$ in
the final expressions for the filling numbers $n_{\alpha\sigma}$
time evolution (\ref{results}). So one should solve self-consistent
system of equations.

In the presence of Coulomb interaction system of equations for pair
correlators can be written in the compact matrix form:

\begin{eqnarray}
i\frac{\partial}{\partial
t}\widehat{G}=[\widehat{G},\widehat{A}+\widehat{C}]+i(\widehat{\Gamma}\widehat{G}+\widehat{G}\widehat{\Gamma})\
\label{system}
\end{eqnarray}

where matrixes $\widehat{A}$, $\widehat{G}$ are determined by
expressions (\ref{matrix}) and (\ref{matrix1}) correspondingly, and
matrixes $\widehat{\Gamma}$ and $\widehat{C}$ can be written as
$||\Gamma||_{ij}=\delta_{i1}\delta_{j1}\gamma$ and
$||C||_{ij}=\delta_{ij}U_{0}G_{jj}$.

The formal solution of the system for pair correlators
(\ref{system}) can be again found with the help of evolution
operator:

\begin{eqnarray}
\widehat{G}(t)=Te^{[i\int_{0}^{t}(\widehat{A}(t^{'})+\widehat{C}(t^{'}))dt^{'}]}\cdot\widehat{G}(0)
T\cdot
e^{[-i\int_{0}^{t}(\widehat{A}(t^{'})+\widehat{C}(t^{'}))dt^{'}]}\nonumber\\
\end{eqnarray}

As initial charge is localized in the QD with number $j$, the
initial conditions are: $n_{i}(0)=n_{j0}\delta_{ij}$, $n_{a}(0)=0$ .
Then one can obtain the expressions:

\begin{eqnarray}
G_{j^{'}j^{'}}(t)&=&\sum_{k,k^{'}}\Omega_{j^{'}k}^{-1}G_{kk^{'}}(0)\Omega_{k^{'}j^{'}}=n_{0j}|\Omega_{jj^{'}}|^{2}\nonumber\\
G_{aa}(t)&=&n_{0j}|\Omega_{aj}|^{2}\ \label{1}
\end{eqnarray}

where evolution operator
$\Omega=Te^{[-i\int_{0}^{t}(\widehat{A}(t^{'})+\widehat{C}(t^{'}))dt^{'}]}$
is considered. So, one can get equations for the matrix elements of
the evolution operator $\Omega$

\begin{eqnarray}
\dot{\Omega}_{aj}&=&i\Delta\cdot\Omega_{aj}+i\sum_{j^{'}}T\cdot\Omega_{j^{'}j}\nonumber\\
\dot{\Omega}_{j^{'}j}&=&iT\cdot\Omega_{aj}+iU_{0}n_{0j}\cdot|\Omega_{j^{'}j}|^{2}\Omega_{j^{'}j}\
\label{3}
\end{eqnarray}

with initial conditions for the functions ${\Omega}_{aj}(0)=0$,
${\Omega}_{jj}(0)=1$, ${\Omega}_{j^{'}j}(0)=0$. If we are interested
in the collective effects connected with the presence of large
number of coupled QDs $N$, the Coulomb interaction between localized
electrons within the initially empty dots can be neglected, because
the filling numbers amplitude is proportional to $1/N^{2}$. So,
taking into account only Coulomb correlations within the dot with
the initial charge, one can simplify the system of equations
(\ref{3}) in the following way:

\begin{eqnarray}
\ddot{\Omega}_{aj}&=&-(N-1)T^{2}\cdot \Omega_{aj}+i(\Delta-\gamma)\cdot\dot{\Omega}_{aj}+iT\cdot\dot{\Omega}_{jj}\nonumber\\
\dot{\Omega}_{jj}&=&iT\cdot
\Omega_{aj}+iU_0n_{0j}\cdot|\Omega_{jj}|^{2}\Omega_{jj}\ \label{5}
\end{eqnarray}

System of equations (\ref{5}) can be easily solved numerically and
consequently one can analyze localized charge relaxation processes.

\section{Calculation results and discussion}

\begin{figure} [h]
\includegraphics[width=65mm]{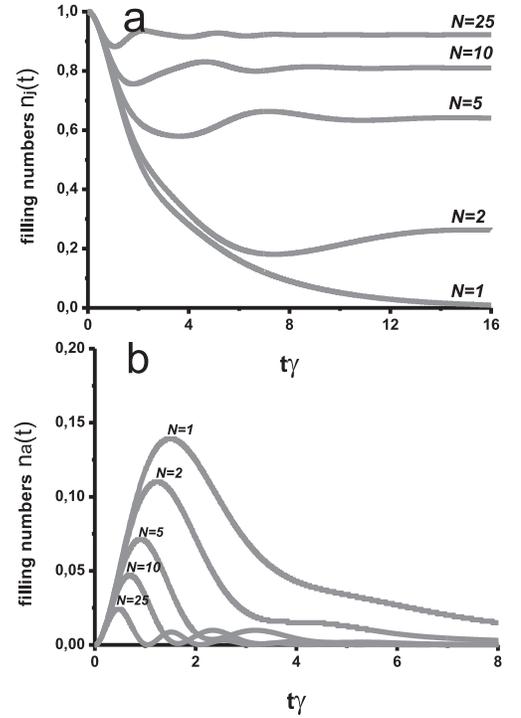}
\caption{Filling numbers time evolution a). in the QD with the
initial charge and b). in the central QD for the different number of
QDs. Parameter $T/\gamma=0.6$ is the same for all the figures.}
\label{figure1a_1b}
\end{figure}

Filling numbers time evolution within the dot with initial charge
$n_{j}(t)$ and within the central QD $n_{a}(t)$ in the absence of
on-site Coulomb repulsion is presented on the Fig.\ref{figure1a_1b}.
The non-resonant tunneling between the dots is considered
($(\varepsilon-\varepsilon_0)/\gamma=-1$).

It is clearly evident that filling numbers relaxation changes
significantly with the increasing of QDs number $N$. When initial
charge is localized in one of the $N$ QDs it remains confined in the
initial dot even in the presence of relaxation processes from the
central dot to the reservoir for the large number of dots. When one
considers two surrounding dots only twenty percent of charge
continue being localized in the initial QD (Fig.\ref{figure1a_1b}a).
But for ten interacting QDs more then eighty percent of charge is
confined in the initial dot (Fig.\ref{figure1a_1b}a). This effect
can be called "charge trapping" and the proposed system of coupled
QDs can be considered as a "charge trap". QDs number $N$ increasing
also leads to the decreasing of charge amplitude in the central QD
$n_{a}(t)$ for a fixed value of ratio $T/\gamma$ due to the
effective growth of tunneling coupling (Fig.\ref{figure1a_1b}b).

\begin{figure*} [t]
\includegraphics[width=140mm]{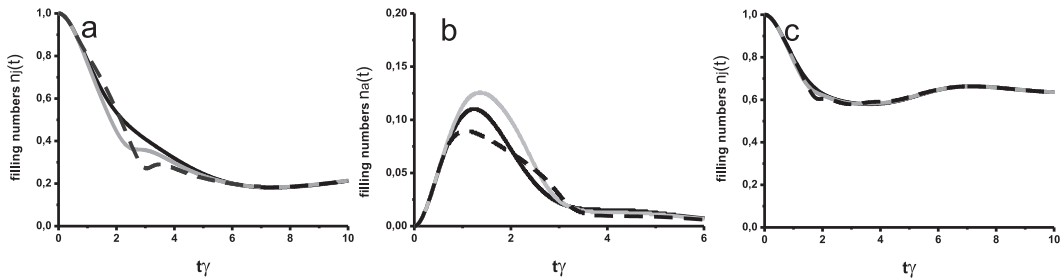}
\caption{Filling numbers time evolution a),c). in the QD with
initial charge and b). in the central QD for the different number of
QDs a),b). $N=2$ and c). $N=5$. Coulomb interaction is taken into
account in the central QD: $U/\gamma=0$-black line,
$U/\gamma=10$-grey line and $U/\gamma=30$-black-dashed line.
Parameter $T/\gamma=0.6$ is the same for all the figures.}
\label{figure2a_2d}
\end{figure*}

Typical calculation results, in the case when on-site Coulomb
repulsion is considered only in the central QD where localized
charge is absent at the initial time moment, are demonstrated on the
Fig.\ref{figure2a_2d}. "Charge trapping" effect is clearly evident
with the increasing of QDs number even in the presence of Coulomb
interaction between localized electrons (Fig.\ref{figure2a_2d}).

For two QDs interacting with the central one Coulomb correlations
strongly influence on the filling numbers relaxation
(Fig.\ref{figure2a_2d}a). A critical value of on-site Coulomb
repulsion exists for a given set of system parameters which
corresponds to the full compensation of the initial negative
detuning \cite{Arseyev_2}. For the smaller values of Coulomb
interaction, correlations lead to the increasing of relaxation rate
in the QD with the initial charge (Fig.\ref{figure2a_2d}a grey line)
in comparison with the case when Coulomb interaction is absent
(Fig.\ref{figure2a_2d}a black line), due to the decreasing of the
initial detuning value. For the values of on-site Coulomb repulsion
larger than the critical one, positive detuning occurs and filling
numbers relaxation rate decreases as a result of positive detuning
value increasing (Fig.\ref{figure2a_2d}a black-dashed line).

\begin{figure*} [t]
\includegraphics[width=140mm]{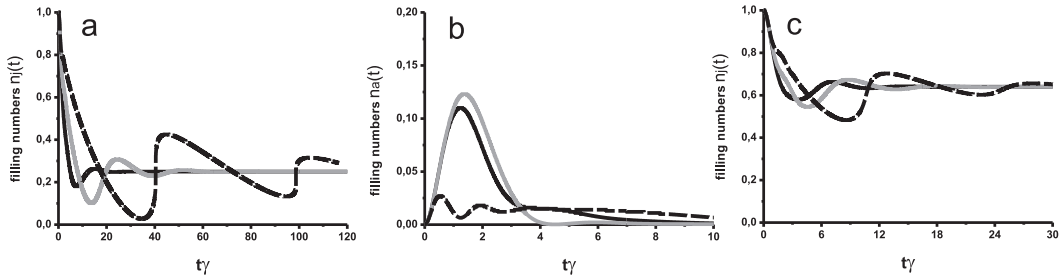}
\caption{Filling numbers time evolution a),c). in the QD with
initial charge and b). in the central QD for the different number of
QDs a),b). $N=2$ and c). $N=5$. Coulomb interaction is taken into
account in the QD with initially localized charge: $U=0$-black line,
$U/\gamma=2$-grey line and $U/\gamma=7$-black-dashed line. Parameter
$T/\gamma=0.6$ is the same for all the figures.} \label{figure3a_3d}
\end{figure*}

With the increasing of QDs number all the effects mentioned above
are still valid, but they are less pronounced
(Fig.\ref{figure2a_2d}c). So the role of Coulomb correlations is
suppressed for the large number of QDs due to the decreasing of
electrons occupation in the central dot.

\begin{figure} [h]
\includegraphics[width=60mm]{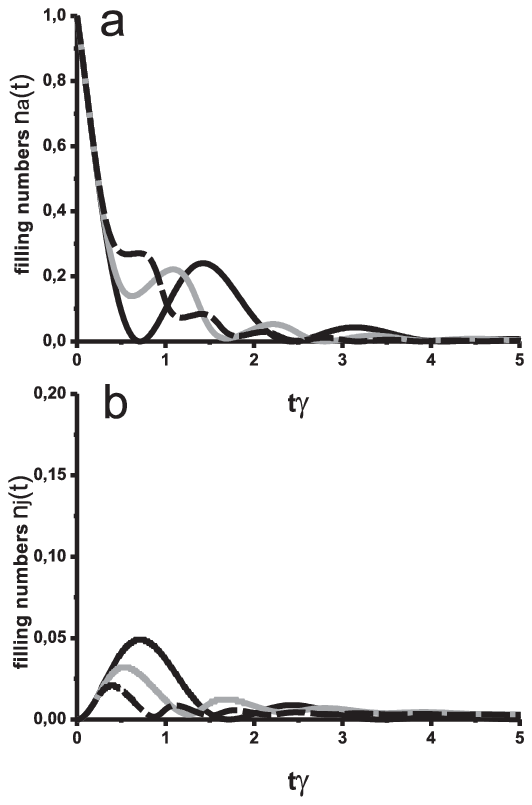}
\caption{Filling numbers time evolution a),c). in the central QD
with initial charge and b),d). in one of the $N$ surrounding QDs for
a),b). $N=2$ and c),d). $N=10$. Coulomb interaction is taken into
account in the central QD: $U/\gamma=0$-black line,
$U/\gamma=4$-grey line and $U/\gamma=8$-black-dashed line. Parameter
$T/\gamma=0.6$ is the same for all the figures.} \label{figure4a_4b}
\end{figure}

Let us now analyze the situation when Coulomb interaction between
localized electrons is taken into account within all the $N$ QDs
which interact with the central one. Calculation results are
presented on the Fig.\ref{figure3a_3d}a and demonstrate "charge
trapping" effect with the increasing of the QDs number. In the case
of two QDs interacting with the central one Coulomb correlations
reveal significantly stronger influence on the filling numbers
relaxation processes in comparison with the geometry when five dots
are considered (Fig.\ref{figure3a_3d}c).

Again two typical relaxation regimes were revealed. The first one
corresponds to the decreasing of initial negative detuning value. In
this regime Coulomb correlations lead to the increasing of
relaxation rate in the QD with initial charge in comparison with the
case when Coulomb interaction is absent (Fig.\ref{figure3a_3d}a grey
and black lines correspondingly). The second one deals with the
Coulomb energy values large enough to compensate negative detuning
and to form the positive one. In this regime filling numbers
relaxation rate decreases as a result of positive detuning value
increasing caused by the Coulomb interaction (Fig.\ref{figure3a_3d}a
grey and black-dashed lines correspondingly).

QDs number increasing also results in the increasing of filling
numbers oscillations frequency. Filling numbers oscillations
frequency for the small Coulomb values decreases corresponding to
the detuning decreasing and increases as a result of positive
detuning formation (Fig.\ref{figure3a_3d}a,c). We found the growth
of charge amplitude in the central QD with the increasing of the
dots number when the negative detuning value decreases and amplitude
decreasing when positive detuning value increases
(Fig.\ref{figure3a_3d}b).

When initial charge is localized in the central QD which is
connected not only to the surrounding QDs but also to the continuous
spectrum states "charge trapping" effect doesn't exist at all
(Fig.\ref{figure4a_4b}).

We now introduce the possible QDs geometry which allows to perform
an experimental observations of charge trapping effects within the
single dot. The most simple configuration is: $N$ similar lateral
QDs interacting only with the single vertically aligned dot. Single
vertical dot is also connected to the continuous spectrum states.
But this geometry reveals a problem of initial charge localization.
Initial charge can be localized in the single QD in the most simple
way by means of the gate voltage. So it is convenient to have a
system with $N-1$ lateral dots and single vertical dot with the
localized charge. These $N$ dots interact only with the single
central vertically aligned QD also connected to the continuous
spectrum states.

To conclude, we have analyzed time evolution of the electron filling
numbers in the system of $N$ interacting QDs both in the absence and
in the presence of Coulomb interaction between localized electrons
within the dots. It was found that Coulomb interaction modifies the
relaxation rates and the character of localized charge time
evolution. It was shown that several time ranges with considerably
different relaxation rates arise in the system of coupled QDs. We
demonstrated and carefully analyzed the presence of strong charge
trapping effects in the proposed systems. It was found that
interacting dots can form an effective high quality charge trap. We
also revealed the Coulomb correlations suppression with the
increasing of QDs number.

The QDs geometry which allows to perform an experimental
observations of charge trapping effects was suggested.

This work was partly supported by the RFBR grants.


\begin{thebibliography}{99}

\bibitem{Wiel}
W. G. van derWiel, S. De Franceschi, J. M. Elzerman et.al., {\it
Rev. Mod. Phys.}, \textbf{75}, 1 (2002).

\bibitem{Potok}
R. M. Potok, I. G. Rau, H. Shtrikman et.al., {\it Nature},
\textbf{446}, 167 (2007).

\bibitem{Hayashi}
T. Hayashi, T. Fujisawa, H. D. Cheong et.al., {\it Phys. Rev.
Lett.}, \textbf{91}, 226804 (2003).

\bibitem{Stafford}
C.A. Stafford, S. Das Sarma, {\it Phys. Rev. Lett.}, \textbf{72},
3590 (1994).

\bibitem{Matveev}
K.A. Matveev, L.I. Glazman, H.U. Baranger, {\it Phys. Rev. B},
\textbf{54}, 5637 (1996).


\bibitem{Boese}
D. Boese, W. Hofstetter, H. Schoeller, {\it Phys. Rev. B},
\textbf{66}, 125315 (2002).

\bibitem{Kikoin0}
K. Kikoin, Y. Avishai, {\it Phys. Rev. B}, \textbf{65}, 115329
(2002).

\bibitem{Orellana}
P.A. Orellana, G.A. Lara, E.V. Anda, {\it Phys. Rev. B},
\textbf{65}, 155317 (2002).

\bibitem{Arseyev_1}
P.I. Arseyev, N.S. Maslova, V.N. Mantsevich, {\it Solid State
Comm.}, \textbf{152}, 1545 (2012).

\bibitem{Arseyev_2}
P.I. Arseyev, N.S. Maslova, V.N. Mantsevich, {\it European Physical
Journal B}, \textbf{85}(7), 249 (2012).


\bibitem{Stafford_1}
C.A. Stafford, N. Wingreen, {\it Phys. Rev. Lett.}, \textbf{76},
1916 (1996).


\bibitem{Hazelzet}
B.L. Hazelzet, M.R. Wegewijs, T. H. Stoof, {\it Phys. Rev. B},
\textbf{63}, 165313 (2001).

\bibitem{Cota}
E. Cota, R. Aguadado, G. Platero, {\it Phys. Rev. Lett.},
\textbf{94}, 107202 (2005).

\bibitem{Contreras-Pulido}
 L.D. Contreras-Pulido,
J. Splettstoesser, M. Governale et.al., {\it Phys. Rev. B},
\textbf{85}, 075301 (2012).

\bibitem{Elste}
Florian Elste, David R. Reichman, and Andrew J. Millis, {\it Phys.
Rev. B}, \textbf{81}, 205413 (2010).

\bibitem{Kennes}
D. M. Kennes, S. G. Jakobs, C. Karrasch et.al., {\it Phys. Rev. B},
\textbf{85}, 085113 (2012).


\bibitem{Kastner}
M. A. Kastner, {\it Rev. Mod. Phys.}, \textbf{64}, 849 (1992).

\bibitem{Beenakker}
C. W. J. Beenakker, {\it Phys. Rev. B}, \textbf{44}, 1646 (1991).

\bibitem{Alhassid}
Y. Alhassid, {\it Rev. Mod. Phys.}, \textbf{72}, 895 (2000).

\bibitem{Vamivakas}
A.N. Vamivakas, C.-Y. Lu, C. Matthiesen et.al., {\it Nature
Letters}, \textbf{467}, 297 (2010).

\bibitem{Stinaff}
E.A. Stinaff, M. Scheibner, A.S. Bracker et.al., {\it Science},
\textbf{311}, 636 (2006).

\bibitem{Kastner_1}
M. A. Kastner, {\it Phys. Today}, \textbf{46}(1), 24 (1993).

\bibitem{Ashoori}
R.C. Ashoori, {\it Nature}, \textbf{379}, 413 (1996).

\bibitem{Chan}
I. Chan, P. Fallahi, A. Vidan et.al., {\it Nanotechnology},
\textbf{15}, 609 (2004).

\bibitem{Kuno_1}
M. Kuno, D.P. Fromm, H.F. Hamann et.al., {\it J. Chem. Phys.},
\textbf{112}, 3117 (2000).

\bibitem{Kuno_2}
M. Kuno, D.P. Fromm, H.F. Hamann et.al., {\it J. Chem. Phys.},
\textbf{115}, 1028 (2001).

\bibitem{Hummon}
M.R. Hummon, A.J. Stollenwerk, V. Narayanamurti, {\it Phys. Rev. B},
\textbf{81}, 115439 (2010).

\bibitem{Brum}
J.A. Brum, P. Hawrylak, {\it Superlattices Microstruct.},
\textbf{22}, 431 (1997).

\bibitem{Pioro}
M. Pioro-Ladriere, M.R. Abolfath, P. Zawadzki et.al., {\it Phys.
Rev. B}, \textbf{72}, 125307 (2005).

\bibitem{Pujari}
Bhalchandra. S. Pujari, Kavita Joshi, D.G. Kanhere et.al., {\it
Phys. Rev. B}, \textbf{78}, 125414 (2008).

\bibitem {Keldysh}
L.V.Keldysh, {\em Sov. Phys JETP}, {\bf 20}, 1018 (1964).

\bibitem{Kikoin}
K. Kikoin, Y. Avishai, {\it Phys. Rev. Lett.}, \textbf{86}, 2090
(2001).

\bibitem{Coleman}
P. Coleman, {\it Phys. Rev. B}, \textbf{29}, 3035 (1984).

\bibitem{Reimann}
S.M. Reimann, M. Manninen, {\it Rev. Mod. Phys.}, \textbf{74}, 1283
(2002).

\bibitem{Foulkes}
W.M.C. Foulkes, L. Mitas, R.J. Needs et.al., {\it Rev. Mod. Phys.},
\textbf{73}, 33 (2001).

\bibitem{Cummings}
F.W. Cummings {\it Phys. Rev. A}, \textbf{33}, 1683 (1986).


\end{thebibliography}
\end{document}